\documentclass[a4paper,graphics,natbib,graphicx,psfig,amssymb,ccaption]{article}
\usepackage{lscape}
\usepackage{graphicx}
\newcommand{\affil}[1]{$^{\rm #1}$}
\usepackage[authoryear]{natbib}
\bibpunct{(}{)}{;}{a}{}{,}
\date{} 

\title{\large\bf\flushleft Colour Transformations between $BVR_c$ and $g'r'i'$ Photometric Systems for Giant Stars}
\author{\parbox{\textwidth}{\flushleft
\vspace{-0.5cm}
{\it S. Ak\affil{A},  T. Ak\affil{A}, S. Karaali\affil{A}, S. Bilir\affil{A},  S. Tun{\c c}el G\"u{\c c}tekin\affil{A}, \"O. \"Onal Ta{\c s}\affil{A}, N. D. \"Ozt\"urkmen\affil{A}, {\c S}. Duran\affil{A}, B. Co{\c s}kuno{\v g}lu\affil{A}, T. Yontan\affil{A}, E. Yaz G{\"o}k{\c c}e\affil{A} and Z. Eker\affil{B}}\\
\vspace{0.4cm}
{\small \affil{A}\,Istanbul University, Faculty of Sciences, Department 
of Astronomy and Space Sciences, 34119 University, Istanbul, Turkey, Email: akserap@istanbul.edu.tr}\\
{\small \affil{B}\,Akdeniz University, Faculty of Sciences, Department of Space Sciences and Technologies, 07058, Antalya, Turkey}\\
}}
\begin{document}
\twocolumn[
\begin{changemargin}{.8cm}{.5cm}
\begin{minipage}{.9\textwidth}
\vspace{-1cm}
\maketitle
%
%
\small{{\bf Abstract:} The transformation equations from $BVR_c$ to $g'r'i'$ magnitudes and vice versa for the giants were established from a sample of 80 stars collected from \citet{Soubiran10} with confirmed surface gravity ($2\leq \log g$ (cms$^{-2}$)$ \leq3$) at effective temperatures $4000<T_{eff} (K)<16000$. The photometric observations, all sample stars at $g'r'i'$ and 65 of them at $BVR_c$, were obtained at T\"UB\.ITAK National Observatory (TUG) 1m (T100) telescope, on the Taurus Mountains in Turkey. The $M_V$ absolute magnitudes of the giant stars were estimated from the absolute magnitude-temperature data for the giant stars by \citet{Sung13} using the $T_{eff}$ from the intrinsic colours considered in this study. The transformation equations could be considered to be valid through the ranges of the following magnitudes and colours involved: $7.10<V_0<14.50$, $7.30<g'_0<14.85$, $-0.20<(B-V)_0<1.41$, $-0.11<(V-R_c)_0<0.73$, $-0.42<(g'-r')_0<1.15$, and $-0.37<(r'-i')_0<0.47$ mag. The transformations were successfully applied to the synthetic $BVR_c$ data of 427 field giants in order to obtain the $g'r'i'$ magnitudes and colours. Comparisons of these data with the $g'r'i'$ observations of giants in this study show that the mean residuals and standard deviations lie within [-0.010, 0.042] and [0.028, 0.068] mag, respectively. 
}
\medskip
\medskip
\end{minipage}
\end{changemargin}
]
\small

\section{Introduction}

All sky surveys have a great impact on our understanding of the Galactic structure. The optical and longer wavelength surveys give detailed information about the Galactic halo, and Galactic disc and bulge, respectively. The Sloan Digital Sky Survey
\citep[SDSS;][]{York00} is one of the most widely used sky. Also, it is the largest photometric and spectroscopic survey in the optical wavelengths. Another widely used sky survey is the Two Micron All Sky Survey \citep[2MASS;][]{Skrutskie06}, which imaged the sky across near-infrared wavelengths. The third sky survey, which is an astrometrically and photometrically important survey, is {\em Hipparcos} \citep{Perryman97}, re-reduced by \citet{vanLeeuwen07}.

SDSS is based on two sets of passbands, i.e. $u'g'r'i'z'$ and $ugriz$. For the first set, the standard Sloan photometric system was defined on the 1m telescope of the USNO Flagstaff Station \citep{Smith02}, while a 2.5m telescope was used for the second passband set \citep{Fukugita96, Gunn98, Hogg01}. The two sets of passbands are very similar, but not quite identical. However, one can use the transformation equations in the literature to make necessary transformations between two systems \citep[cf.][]{Rider04}.

It has been customary to derive transformation equations between a newly defined photometric system and the traditional ones, such as the Johnson-Cousins $UBVR_cI_c$ system. Several transformations can be found in the literature related SDSS photometric system \citep*{Smith02,Karaali05,Bilir05,Rodgers06,Jordi06,Chonis08,Bilir08}. All these transformations are devoted to dwarfs, the most populated luminosity class in our Galaxy. The two transformations which are carried out for red giants are those of \citet{Yaz10} and \citet{Karaali13}.

\citet{Yaz10} used the $JHK_s$, $BVI$, and $gri$ magnitudes in the 2MASS \citep{Cutri03}, reduced {\em Hipparcos'} \citep{vanLeeuwen07} and Ofek's (\citeyear{Ofek08}) catalogues and identified two samples of red giants by matching them with the Cayrel de Strobel et al.'s (\citeyear{Cayrel01}) spectroscopic catalogue which contains the surface gravity $\log g$, a parameter available for dwarf-giant separation. The first sample of stars (91 giants) was used for the transformations between $JHK_s$ and $BVI$, while the second one (82 giants) was devoted to transformations between $JHK_s$ and $gri$. The transformations of \citet{Karaali13} are based on synthetic $UBV$ colours of \citet{Buser92} and synthetic $ugr$ colours of \citet{Lenz98}. They are metallicity and two colours dependent. Three sets of transformation equations are obtained, i.e. for $[M/H]=0, -1, -2$ dex, which can be interpolated/extrapolated to different metallicities. The advantage of these transformations is that they can be used to extend the colour ranges of the observed $u-g$ and $g-r$ colours which are restricted due to the saturation of the SDSS magnitudes.    

In this study, we present transformation equations between either of the most widely used sky surveys, SDSS $g'r'i'$, and $BVR_c$ for giants. The equations are based on the data which were observed and reduced by a team who are included as co-authors in this paper. The sections are organised as follows. Data are presented in Section 2. Section 3 is devoted to the transformation equations and their application and a summary and discussions are given in Section 4.

\section{The data}

\subsection{Observations}
The sample consists of fairly bright 80 giants identified and collected by their surface gravities and effective temperatures in the Pastel catalogue \citep{Soubiran10}: i) $2\leq \log g$~(cms$^{-2}$) $\leq 3$, ii) $4000<T_{eff}(K)<16000$. The sample stars cover a large range of metallicities, i.e. $-4\leq [Fe/H]\leq 0.5$ dex. The sample stars were observed at least three exposures in each filter used. The observations were carried out at 1m RC telescope (T100) at T\"UB\.ITAK National Observatory (TUG)\footnote{www.tug.tubitak.gov.tr} at Bak{\i}rl{\i}tepe, Antalya, in Turkey from 2011 July through 2012 September. Table 1 summarizes the journal of observations. The columns indicate the dates, the number of nights, and the number of observing sets at $g'r'i'$ and $BVR_c$ filters both for the sample and standard stars.

Standard reduction techniques were performed with Image Reduction and Analysis Facility (IRAF)\footnote{IRAF is distributed by the National Optical Astronomy Observatories}. Sky observations during the twilight were used in the flat field corrections. As the dark current level of the CCD camera is very low (0.0002 e$^{-}$~pixel$^{-1}$~s$^{-1}$) and the exposure times during observing runs are not long, dark current corrections were not applied. Instrumental magnitudes were obtained through aperture photometry using standard IRAF software packages. The instrumental magnitudes and colours were transformed to the standard photometric systems following the procedures described below. The following equations were used for the transformation of the instrumental magnitudes and colours of giant stars to the standard magnitudes and colours:

\begin{equation}
V = v-k_V\times X_v +\epsilon_V \times (B-V)+\zeta_V
\end{equation}

\begin{equation}
B-V=\mu \times[(b-v)-(k_{BV} \times X_{BV})]+\zeta_{BV}
\end{equation}

\begin{equation}
V-R_c=\rho \times[(v-r_c)-(k_{VR_c} \times X_{VR_c})]+\zeta_{VR_c}
\end{equation}

\begin{equation}
g'=g-k_g \times X_g+\epsilon_g \times(g-r)+\zeta_{g} 
\end{equation}

\begin{equation}
g'-r'=\kappa \times[(g-r)-(k_{gr} \times X_{gr})]+\zeta_{gr}
\end{equation}

\begin{equation}
r'-i'=\tau \times[(r-i)-(k_{ri}\times X_{ri})]+\zeta_{ri}
\end{equation}
where, $b$, $v$, $r_c$ and $B$, $V$, $R_c$ are the instrumental and the standard magnitudes, respectively. Similarly, $g$, $r$, $i$ and $g'$, $r'$, $i'$ denote the instrumental and standard magnitudes for SDSS $u'g'r'i'z'$ photometric system. $k$ and $X$ are photometric extinction coefficient and airmass, respectively, with a subscript denoting the filter or the colour. $\epsilon$, $\mu$, $\rho$, $\kappa$ and $\tau$ are the transformation coefficients from the instrumental to the standard, where $\zeta$ denotes nightly photometric zeropoint with a subscript indicating the filter or the colour. Using the instrumental magnitudes of the standard stars 
\citep{Landolt09} measured through observations and applying multiple linear least square fits to the equations above, the photometric extinction coefficients, the transformation coefficients and the zeropoint constants were estimated on each night. The averages of them are listed in Table 2. The final standardized photometric data of the sample stars are listed in Table 3. The $B$, $V$, $R_c$ observations exist only for 65 sample stars. Therefore, we had to use the synthetic $B$, $V$, $R_c$ magnitudes in \citet{Pickles10} to estimate intrinsic $B-V$ and $V-R_c$ colours for the 15 stars  which are not observed at $BVR_c$ but observed at $g'r'i'$.   

\begin{table}
\setlength{\tabcolsep}{5pt}
{\scriptsize
\center
\caption{Observing runs at the TUG T100 telescope. Number of frames for giants and standard stars are given in the last four columns according to the filter sets used.}
\begin{tabular}{rccccc}
\hline
           &            & \multicolumn{2}{c}{Giants} & \multicolumn{2}{c}{Std. stars} \\
\hline
\multicolumn{1}{c}{Date}      &     Nights &   $g'r'i'$ &     $BVR_c$&   $g'r'i'$ &        $BVR_c$ \\
\hline
July 21-23, 2011 &          3 &        137 &        139 &        187 &        185 \\
Sept. 9-11, 2011 &          3 &        210 &        187 &        207 &        172 \\
Oct. 31-Nov. 01, 2011 &     2 &         74 &         62 &         91 &         43 \\
Nov. 21-22, 2011 &          2 &         60 &         45 &         74 &         39 \\
Dec. 29,    2011 &          1 &         68 &        --- &        110 &        --- \\
Feb. 19,    2012 &          1 &        162 &        --- &         79 &        --- \\
Mar. 29,    2012 &          1 &        137 &        --- &        169 &        --- \\
Apr. 01,    2012 &          1 &        --- &         57 &        --- &         60 \\
Apr. 12,    2012 &          1 &        --- &         56 &        --- &         37 \\
Apr. 16-17, 2012 &          2 &         88 &         87 &         63 &         56 \\
July 28-29, 2012 &          2 &         99 &         66 &        203 &        141 \\
Aug. 18-19, 2012 &          2 &        126 &        121 &        177 &        111 \\
Sep. 16-18, 2012 &          3 &        248 &        368 &        203 &        203 \\
\hline
     Total &         24 &       1409 &       1188 &       1563 &       1047 \\
\hline
\end{tabular}  
}
\end{table}

\begin{table}
\setlength{\tabcolsep}{3pt}
{\scriptsize
\center
\caption{Derived average photometric extinction coefficients ($k$), transformation coefficients ($C$, see the text), zeropoints ($\zeta$) and average standard deviation ($\sigma$) of fits for observing runs. Error values show standard deviation of the measurements.}
\begin{tabular}{lcccc}
\hline
Magnitude/ & $k$ & $C$ & $\zeta$ &  $\sigma$ \\
colour &  &  &  &  \\
\hline
$V$		& 0.137$\pm$0.022 & -0.076$\pm$0.008 & -0.753$\pm$0.099 & 0.015$\pm$0.007 \\
$B-V$ 		& 0.082$\pm$0.013 &  1.191$\pm$0.016 & -0.012$\pm$0.035 & 0.010$\pm$0.003 \\
$V-R_c$		& 0.044$\pm$0.023 &  0.942$\pm$0.009 &  0.068$\pm$0.021 & 0.008$\pm$0.002 \\
$g'$		& 0.179$\pm$0.048 &  0.021$\pm$0.005 & -0.229$\pm$0.092 & 0.016$\pm$0.007 \\
$g'-r'$		& 0.072$\pm$0.025 &  1.035$\pm$0.009 & -0.247$\pm$0.030 & 0.009$\pm$0.003 \\
$r'-i'$		& 0.058$\pm$0.020 &  0.962$\pm$0.023 & -0.628$\pm$0.029 & 0.010$\pm$0.003 \\
\hline
\end{tabular}
}
\end{table}

\begin{landscape}
\textwidth = 700 pt
\begin{table*}
\setlength{\tabcolsep}{.5pt}
{\tiny
\center
\caption{Photometric data of the sample stars. The columns give: (1) Current number, (2) Star name, (3), (4), (5), and (6) the Equatorial and Galactic coordinates, (7) colour excess of the star, (8)-(13) magnitudes, colours and their errors in $BVR_c$ photometric system, (14) References for the columns (8)-(13), (15)-(20) magnitudes, colours and their errors in $g'r'i'$ photometric system, (21) effective temperature $T_{eff}$, (22) surface gravity $\log g$, (23) the metallicity $[Fe/H]$ and (24) references for the columns (21)-(23).}   
\begin{tabular}{clcccccccccccccccccccccc}
\hline
(1) & (2) & (3) & (4) & (5)& (6)& (7)& (8)& (9) & (10)&  (11) & (12) & (13) & (14) & (15) & (16) & (17) & (18) & (19) & (20) & (21) & (22) & (23) & (24)\\
	ID &       Star &          $\alpha$  (J2000) &          $\delta$ (J2000) &          $l$  &          $b$ &    $E_{d}(B-V)$ &          $V$ &       $V_{err}$ &      $(B-V)$ &   $(B-V)_{err}$ &        $V-R_c$ &   $(V-R_c)_{err}$ &        Ref &
         $g'$ &     $g'_{err}$ &      $(g'-r')$ &  $(g'-r')_{err}$ &      $(r'-i')$ &  $(r'-i')_{err}$ &       $T_{eff}$ &       $\log g$  & $[Fe/H]$ & Ref \\

	&       &          (hh:mm:ss.ss) &          (dd:mm:ss.ss) &          (deg) &          (deg)&    (mag) &          (mag)  &       (mag) &      (mag) &   (mag) &     (mag) &   (mag) &        &
         (mag) &     (mag) &      (mag) &  (mag) &      (mag) &  (mag) &       (K) &       (cms$^{-2}$) &(dex) &      \\
\hline
         1 &      HD 26    & 00 05 22.20 & +08 47 16.10 &    104.069 &    -52.389 &      0.062 &      8.186 &      0.002 &      1.066 &      0.003 &      0.507 &      0.003 &          1 &      8.646 &      0.002 &      0.740 &      0.003 &      0.255 &      0.003 &       5250 &       2.59 &      -0.29 &1999ApJ...521..753T \\
         2 & BD+28 0054    & 00 23 43.09 & +29 24 03.62 &    115.725 &    -33.083 &      0.036 &      9.890 &      0.002 &      0.535 &      0.003 &      0.304 &      0.003 &          1 &     10.037 &      0.002 &      0.195 &      0.003 &     -0.047 &      0.004 &       6337 &       2.84 &      -0.06 &1995AJ....110.2319C \\
         3 & BD+71 0031    & 00 43 44.35 & +72 10 43.12 &    122.335 &      9.316 &      0.360 &     10.101 &        --- &      0.469 &        --- &      0.291 &        --- &          2 &     10.344 &      0.001 &      0.249 &      0.001 &      0.085 &      0.002 &       6129 &       3.75 &      -1.81 &2008MNRAS.391...95R \\
         4 &    HD 6755    & 01 09 43.06 & +61 32 50.19 &    125.108 &     -1.247 &      0.032 &      7.731 &      0.002 &      0.731 &      0.003 &      0.431 &      0.003 &          1 &      7.985 &      0.002 &      0.534 &      0.003 &      0.202 &      0.003 &       5100 &       2.70 &     -1.47 &2001A\&A...370..951M \\
         5 &    HD 9356    & 01 32 08.17 & +01 20 30.23 &    143.544 &    -59.891 &      0.021 &      9.823 &      0.001 &      0.413 &      0.001 &      0.265 &      0.002 &          1 &     10.027 &      0.002 &      0.178 &      0.003 &     -0.001 &      0.003 &       6282 &       2.77 &      -1.38 &1995AJ....110.2319C \\
         6 &   HD 19510    & 03 08 30.88 & +10 26 45.22 &    169.083 &    -39.838 &      0.174 &      9.831 &      0.002 &      0.582 &      0.003 &      0.379 &      0.003 &          1 &     10.080 &      0.002 &      0.447 &      0.004 &      0.156 &      0.004 &       6109 &       2.60 &      -2.50 &1995AJ....110.2319C \\
         7 & BD+26 0595    & 03 40 10.04 & +26 57 37.21 &    163.275 &    -22.401 &      0.160 &      8.315 &      0.002 &      1.060 &      0.003 &      0.578 &      0.003 &          1 &      8.671 &      0.001 &      0.868 &      0.001 &      0.359 &      0.001 &       4383 &       2.10 &       -0.80 &1984Thesis. Proust \\
         8 &  HD 281679    & 04 09 16.98 & +30 46 33.47 &    165.534 &    -15.288 &      0.490 &      9.322 &      0.002 &      0.577 &      0.003 &      0.285 &      0.003 &          1 &      9.582 &      0.002 &      0.335 &      0.002 &      0.152 &      0.002 &       8542 &       2.50 &      -1.43 &1987SvA....31...37K \\
         9 &   HD 26886    & 04 14 58.83 & -00 59 50.82 &    193.602 &    -34.703 &      0.074 &      7.878 &      0.001 &      0.985 &      0.002 &      0.497 &      0.001 &          1 &      8.301 &      0.002 &      0.700 &      0.002 &      0.255 &      0.002 &       4802 &       2.22 &      -0.28 &2003A\&A...257..265L \\
        10 &   HD 27271    & 04 18 33.82 & +02 28 13.92 &    190.740 &    -32.035 &      0.075 &      7.471 &      0.001 &      0.969 &      0.002 &      0.474 &      0.001 &          1 &      7.980 &      0.001 &      0.722 &      0.001 &      0.255 &      0.001 &       4874 &       2.98 &      -0.06 &2003A\&A...257..265L \\
        11 & NGC 2112 204  & 05 53 47.54 & +00 22 02.00 &    205.916 &    -12.625 &      0.793 &     11.822 &      0.002 &      1.786 &      0.004 &      1.000 &      0.003 &          1 &     12.937 &      0.003 &      1.571 &      0.004 &      0.749 &      0.003 &       4550 &       2.46 &      -0.17 &1996AJ....112.1551B \\
        12 & NGC 2112 402  & 05 53 53.30 & +00 24 37.90 &    205.889 &    -12.584 &      0.839 &     11.512 &      0.001 &      1.469 &      0.003 &      0.818 &      0.002 &          1 &     12.423 &      0.002 &      1.252 &      0.003 &      0.593 &      0.003 &       5050 &       2.72 &      -0.05 &1996AJ....112.1551B \\
        13 &  HD 252940    & 06 11 37.25 & +26 27 30.11 &    185.021 &      3.752 &      0.133 &      9.110 &      0.002 &      0.206 &      0.003 &      0.148 &      0.004 &          1 &      9.264 &      0.002 &      0.107 &      0.003 &     -0.068 &      0.003 &       7550 &       2.95 &      -1.77 &2000A\&A...364..102K \\
        14 & BD+37 1458    & 06 16 01.52 & +37 43 18.76 &    175.418 &      9.825 &      0.105 &      8.857 &      0.002 &      0.684 &      0.003 &      0.408 &      0.003 &          1 &      9.150 &      0.001 &      0.430 &      0.002 &      0.193 &      0.003 &       5100 &       2.90 &      -2.31 &1999AJ....118..527F \\
        15 & BD+38 1456    & 06 20 03.84 & +38 20 44.11 &    175.211 &     10.815 &      0.142 &     10.697 &      0.001 &      0.897 &      0.001 &      0.457 &      0.001 &          1 &     11.203 &      0.002 &      0.638 &      0.003 &      0.260 &      0.003 &       5000 &       2.75 &      -1.50 &2003PASP..115...22Y \\
        16 &   HD 58337    & 07 26 01.95 & +21 54 46.00 &    196.614 &     17.118 &      0.033 &      9.700 &      0.001 &      1.323 &      0.002 &      0.605 &      0.001 &          1 &     10.347 &      0.002 &      0.964 &      0.003 &      0.352 &      0.003 &       4582 &       2.10 &      -0.40 &1984ApJS...55...27D \\
        17 & NGC 2420 115  & 07 38 21.67 & +21 33 51.40 &    198.114 &     19.626 &      0.036 &     11.591 &      0.003 &      1.327 &      0.007 &      0.647 &      0.004 &          1 &     12.068 &      0.002 &      0.849 &      0.003 &      0.273 &      0.003 &       4541 &       2.20 &      -0.60 &1987AJ.....93..359S \\
        18 & NGC 2420 173  & 07 38 26.96 & +21 33 31.30 &    198.128 &     19.643 &      0.036 &     11.807 &      0.003 &      1.095 &      0.007 &      0.528 &      0.004 &          1 &     12.161 &      0.002 &      0.629 &      0.003 &      0.211 &      0.003 &       4893 &       2.50 &      -0.55 &1987AJ.....93..359S \\
        19 & BD-01 1792    & 07 39 50.11 & -01 31 20.37 &    219.884 &     10.042 &      0.041 &      9.122 &      0.002 &      0.938 &      0.003 &      0.433 &      0.002 &          1 &      9.628 &      0.002 &      0.645 &      0.003 &      0.203 &      0.003 &       4850 &       2.70 &      -1.26 &2000AJ....120.1841F \\
        20 & BD+80 0245    & 08 11 06.23 & +79 54 29.56 &    133.914 &     30.136 &      0.021 &      9.906 &        --- &      0.634 &        --- &      0.376 &        --- &          2 &     10.240 &      0.001 &      0.404 &      0.001 &      0.148 &      0.002 &       5225 &       3.00 &      -2.05 &2000AJ....120.1841F \\
        21 & BD+00 2245    & 08 16 57.77 & +00 01 03.72 &    223.038 &     18.942 &      0.033 &      9.604 &      0.001 &      0.676 &      0.002 &      0.335 &      0.001 &          1 &      9.996 &      0.001 &      0.381 &      0.001 &      0.129 &      0.002 &       5425 &       3.00 &      -1.28 &1998ApJ...500..398R \\
        22 &  HD 233517    & 08 22 46.71 & +53 04 49.19 &    165.382 &     34.891 &      0.044 &      9.675 &        --- &      1.325 &        --- &      0.665 &        --- &          2 &     10.335 &      0.002 &      1.049 &      0.002 &      0.411 &      0.001 &       4475 &       2.25 &      -0.37 &2000ApJ...542..978B \\
        23 & NGC 2682 84   & 08 51 12.70 & +11 52 42.40 &    215.601 &     31.909 &      0.030 &     10.560 &      0.003 &      1.058 &      0.005 &      0.512 &      0.004 &          1 &     11.114 &      0.004 &      0.891 &      0.005 &      0.330 &      0.004 &       4750 &       2.40 &      -0.02 &2000A\&A...360..499T \\
        24 & NGC 2682 105  & 08 51 17.10 & +11 48 16.10 &    215.689 &     31.895 &      0.030 &     10.347 &      0.003 &      1.170 &      0.005 &      0.595 &      0.004 &          1 &     11.023 &      0.003 &      1.019 &      0.004 &      0.368 &      0.004 &       4400 &       2.00 &       0.01 &2010AJ....139.1942F \\
        25 & BD+12 1924    & 08 51 20.10 & +12 18 10.43 &    215.161 &     32.113 &      0.026 &      9.338 &        --- &      1.432 &        --- &      0.738 &        --- &          2 &     10.041 &      0.002 &      1.170 &      0.002 &      0.480 &      0.001 &       4131 &       2.30 &       0.00 &1981A\&A....99..221F \\
        26 & NGC 2682 141  & 08 51 22.80 & +11 48 01.70 &    215.705 &     31.914 &      0.030 &     10.529 &      0.003 &      0.989 &      0.005 &      0.567 &      0.004 &          1 &     11.046 &      0.002 &      0.831 &      0.003 &      0.283 &      0.003 &       4650 &       2.80 &       0.06 &2010A\&A...511A..56P \\
        27 & NGC 2682 151  & 08 51 26.19 & +11 53 52.00 &    215.608 &     31.968 &      0.031 &     10.555 &      0.003 &      0.958 &      0.005 &      0.601 &      0.004 &          1 &     11.006 &      0.002 &      0.822 &      0.003 &      0.280 &      0.003 &       4760 &       2.40 &       0.01 &2000A\&A...360..499T \\
        28 & NGC 2682 164  & 08 51 28.99 & +11 50 33.10 &    215.673 &     31.955 &      0.031 &     10.603 &      0.003 &      0.969 &      0.006 &      0.579 &      0.004 &          1 &     11.117 &      0.002 &      0.855 &      0.003 &      0.293 &      0.003 &       4659 &       2.53 &      -0.02 &2009A\&A...493..309S \\
        29 & NGC 2682 224  & 08 51 43.55 & +11 44 26.40 &    215.811 &     31.966 &      0.025 &     10.875 &      0.004 &      0.965 &      0.006 &      0.600 &      0.004 &          1 &     11.288 &      0.002 &      0.852 &      0.003 &      0.302 &      0.003 &       4710 &       2.40 &      -0.11 &2000A\&A...360..499T \\
        30 & NGC 2682 231  & 08 51 45.08 & +11 47 45.90 &    215.755 &     31.995 &      0.027 &     11.523 &      0.006 &      0.990 &      0.011 &      0.504 &      0.007 &          1 &     11.979 &      0.003 &      0.778 &      0.004 &      0.281 &      0.004 &       4893 &       3.00 &      -0.35 &1980ApJ...241..981C \\
        31 & NGC 2682 266  & 08 51 59.52 & +11 55 04.90 &    215.653 &     32.099 &      0.032 &     10.451 &      0.003 &      1.105 &      0.005 &      0.483 &      0.004 &          1 &     11.007 &      0.002 &      0.825 &      0.003 &      0.287 &      0.003 &       4730 &       2.40 &      -0.02 &2000A\&A...360..499T \\
        32 & BD+23 2130    & 09 39 39.17 & +22 52 15.42 &    207.548 &     46.593 &      0.028 &      9.716 &        --- &      1.104 &        --- &      0.686 &        --- &          2 &     10.212 &      0.002 &      0.858 &      0.002 &      0.284 &      0.002 &       5228 &       2.94 &     -2.42 &2005MNRAS.364..712Z \\
        33 &  HD 233666    & 09 42 19.47 & +53 28 26.15 &    162.399 &     46.535 &      0.008 &      9.225 &        --- &      0.665 &        --- &      0.395 &        --- &          2 &      9.606 &      0.002 &      0.474 &      0.003 &      0.196 &      0.003 &       5300 &       2.50 &      -1.65 &2000ApJ...544..302B \\
        34 &  HD 237846    & 09 52 38.68 & +57 54 58.59 &    155.634 &     46.250 &      0.009 &      9.844 &        --- &      0.816 &        --- &      0.474 &        --- &          2 &     10.267 &      0.001 &      0.541 &      0.002 &      0.231 &      0.003 &       5015 &       2.02 &      -2.81 &2012ApJ...753...64I \\
        35 & BD+10 2179    & 10 38 55.23 & +10 03 48.50 &    235.211 &     54.442 &      0.023 &      9.993 &      0.002 &     -0.160 &      0.002 &     -0.090 &      0.003 &          1 &      9.769 &      0.002 &     -0.390 &      0.003 &     -0.353 &      0.004 &      15750 &       2.80 &        1.40 &1969ApJ...157..721H \\
        36 & BD+09 2384    & 10 40 25.20 & +08 54 03.95 &    237.215 &     54.098 &      0.024 &      9.855 &        --- &      0.885 &        --- &      0.495 &        --- &          2 &     10.256 &      0.002 &      0.624 &      0.003 &      0.226 &      0.003 &       5200 &       3.00 &      -0.71 &1991ApJS...77..515L \\
        37 &    G146 76    & 10 59 57.47 & +44 46 43.75 &    167.171 &     61.600 &      0.009 &     10.422 &        --- &      0.778 &        --- &      0.472 &        --- &          2 &     10.769 &      0.002 &      0.497 &      0.003 &      0.201 &      0.003 &       5202 &       2.85 &      -1.64 &2005MNRAS.364..712Z \\
        38 & BD+32 2188    & 11 47 00.50 & +31 50 08.65 &    190.506 &     75.227 &      0.019 &     10.708 &      0.003 &     -0.032 &      0.004 &     -0.028 &      0.004 &          1 &     10.620 &      0.002 &     -0.264 &      0.003 &     -0.269 &      0.004 &      10450 &       2.10 &       -1.11 &2000A\&A...364..102K \\
        39 & BD+27 2057    & 11 47 28.72 & +26 24 45.56 &    212.163 &     75.716 &      0.019 &      9.550 &      0.001 &      0.872 &      0.001 &      0.431 &      0.001 &          1 &      9.863 &      0.001 &      0.615 &      0.001 &      0.216 &      0.001 &       4810 &       2.25 &      -0.51 &2012AJ....144...20A \\
        40 & BD-01 2582    & 11 53 37.32 & -02 00 36.74 &    275.101 &     57.703 &      0.019 &      9.538 &        --- &      0.782 &        --- &      0.478 &        --- &          2 &      9.899 &      0.001 &      0.498 &      0.001 &      0.206 &      0.001 &       5237 &       2.79 &      -2.14 &2011ApJ...737....9R \\
        41 & BD+29 2231    & 11 55 52.34 & +28 26 14.77 &    203.631 &     77.648 &      0.024 &      9.875 &        --- &      0.965 &        --- &      0.525 &        --- &          2 &     10.252 &      0.002 &      0.675 &      0.003 &      0.234 &      0.003 &       5060 &       2.50 &      -0.39 &2001A\&A...380..578T \\
        42 & BD+25 2436    & 11 56 28.83 & +24 59 16.15 &    219.693 &     77.480 &      0.016 &      9.909 &      0.002 &      0.882 &      0.003 &      0.484 &      0.003 &          1 &     10.323 &      0.002 &      0.678 &      0.003 &      0.237 &      0.003 &       4990 &       2.40 &      -0.48 &2001A\&A...380..578T \\
        43 &  HD 104783    & 12 04 05.21 & +37 59 58.95 &    162.411 &     75.271 &      0.015 &      9.094 &        --- &      0.836 &        --- &      0.474 &        --- &          2 &      9.533 &      0.002 &      0.605 &      0.002 &      0.236 &      0.001 &       5140 &       2.40 &      -0.55 &2001A\&A...380..578T \\
        44 & BD+25 2459    & 12 06 43.44 & +24 42 59.23 &    223.929 &     79.665 &      0.022 &      9.555 &        --- &      0.955 &        --- &      0.505 &        --- &          2 &     10.035 &      0.001 &      0.686 &      0.001 &      0.254 &      0.001 &       4980 &       2.50 &      -0.35 &2001A\&A...380..578T \\
        45 & BD-04 3208    & 12 07 15.07 & -05 44 01.61 &    283.291 &     55.444 &      0.042 &      9.991 &      0.003 &      0.412 &      0.004 &      0.254 &      0.004 &          1 &     10.118 &      0.002 &      0.242 &      0.003 &      0.060 &      0.003 &       5900 &       3.00 &      -2.62 &2000AJ....120.1841F \\
        46 &  HD 105944    & 12 11 29.35 & +44 15 02.56 &    145.469 &     71.088 &      0.012 &      9.889 &      0.002 &      0.863 &      0.003 &      0.461 &      0.002 &          1 &     10.297 &      0.002 &      0.625 &      0.002 &      0.242 &      0.002 &       5090 &       2.10 &      -0.37 &2001A\&A...380..578T \\
        47 & BD+17 2473    & 12 23 31.02 & +16 54 09.24 &    269.223 &     77.906 &      0.023 &     10.083 &      0.002 &      0.759 &      0.003 &      0.424 &      0.003 &          1 &     10.478 &      0.002 &      0.514 &      0.002 &      0.221 &      0.001 &       5050 &       3.00 &      -1.21 &1998ApJ...500..398R \\
        48 & BD+25 2502    & 12 24 17.14 & +24 19 28.31 &    236.060 &     83.274 &      0.018 &      9.922 &      0.003 &      0.786 &      0.004 &      0.438 &      0.004 &          1 &     10.299 &      0.002 &      0.575 &      0.002 &      0.216 &      0.002 &       5090 &       2.20 &      -0.74 &2001A\&A...380..578T \\
        49 & BPS BS 16076-006 & 12 48 22.75 & +20 56 44.04 & 296.328 &     83.778 &      0.028 &     13.539 &      0.004 &      0.590 &      0.006 &      0.386 &      0.005 &          1 &     13.801 &      0.003 &      0.437 &      0.004 &      0.181 &      0.004 &       5199 &       3.00 &      -3.81 &2009A\&A...501..519B \\
        50 & BD+34 2371    & 12 57 49.30 & +33 39 01.13 &    111.410 &     83.335 &      0.013 &      9.451 &        --- &      0.959 &        --- &      0.511 &        --- &          2 &      9.914 &      0.002 &      0.679 &      0.002 &      0.209 &      0.002 &       4980 &       2.50 &      -0.18 &2001A\&A...380..578T \\
        51 &  HD 113321    & 13 02 46.53 & +16 27 59.36 &    317.348 &     79.020 &      0.024 &      9.364 &        --- &      1.026 &        --- &      0.544 &        --- &          2 &      9.838 &      0.002 &      0.776 &      0.002 &      0.269 &      0.001 &       4739 &       2.10 &      -0.07 &2006A\&A...456.1109M \\
        52 & BPS BS 16929-005 & 13 03 29.47 & +33 51 09.14 & 102.568 &     82.793 &      0.011 &     13.642 &      0.004 &      0.642 &      0.006 &      0.423 &      0.006 &          1 &     13.930 &      0.003 &      0.479 &      0.004 &      0.202 &      0.004 &       5245 &       2.70 &      -3.07 &2008ApJ...681.1524L \\
        53 & BD+18 2890    & 14 32 13.48 & +17 25 24.28 &     15.547 &     64.809 &      0.018 &      9.880 &      0.002 &      0.720 &      0.002 &      0.451 &      0.002 &          1 &     10.121 &      0.001 &      0.538 &      0.001 &      0.223 &      0.001 &       5000 &       2.20 &      -1.58 &2000ApJ...544..302B \\
        54 &  HD 128188    & 14 35 46.82 & -11 24 12.26 &    339.673 &     43.900 &      0.099 &     10.119 &      0.002 &      0.947 &      0.003 &      0.592 &      0.002 &          1 &     10.451 &      0.002 &      0.744 &      0.002 &      0.360 &      0.001 &       4657 &       2.03 &      -1.35 &2008A\&A...484L..21M \\
        55 & BPS CS 30312-059 & 15 34 48.84 & -01 23 37.29 &   3.659 &     41.451 &      0.107 &     13.118 &      0.005 &      0.824 &      0.008 &      0.504 &      0.006 &          1 &     13.529 &      0.003 &      0.653 &      0.004 &      0.313 &      0.004 &       5021 &       2.06 &      -3.14 &2008ApJ...681.1524L \\
        56 & BD+11 2998    & 16 30 16.78 & +10 59 51.74 &     26.608 &     36.278 &      0.043 &      9.086 &      0.001 &      0.688 &      0.001 &      0.414 &      0.001 &          1 &      9.359 &      0.002 &      0.516 &      0.002 &      0.196 &      0.002 &       5350 &       2.00 &      -1.38 &1992AJ....104..645K \\
        57 &  HD 156074    & 17 13 31.24 & +42 06 22.76 &     66.998 &     35.404 &      0.012 &      7.617 &      0.001 &      1.195 &      0.001 &      0.539 &      0.001 &          1 &      8.115 &      0.001 &      0.815 &      0.001 &      0.168 &      0.002 &       4755 &       2.05 &      -0.10 &1973A\&A....22..293G \\
        58 & NGC 6341 9012 & 17 16 52.84 & +43 03 29.50 &     68.238 &     34.894 &      0.020 &     14.541 &      0.008 &      0.565 &      0.013 &      0.385 &      0.010 &          1 &     14.837 &      0.008 &      0.457 &      0.010 &      0.171 &      0.008 &       5500 &       3.00 &      -2.34 &2000AJ....120.1351S \\
        59 & NGC 6341 10065 & 17 17 11.41 & +43 06 02.70 &    68.297 &     34.843 &      0.019 &     14.362 &      0.010 &      0.656 &      0.016 &      0.446 &      0.012 &          1 &     14.744 &      0.009 &      0.558 &      0.011 &      0.226 &      0.008 &       5260 &       2.40 &      -2.34 &2000AJ....120.1351S \\
        60 & NGC 6341 11027 & 17 17 21.61 & +43 06 15.90 &    68.306 &     34.813 &      0.019 &     14.519 &      0.007 &      0.684 &      0.012 &      0.452 &      0.009 &          1 &     14.917 &      0.010 &      0.574 &      0.012 &      0.226 &      0.010 &       5150 &       2.20 &      -2.34 &2000AJ....120.1351S \\
\hline
\end{tabular}
(1) This study, (2) \citet{Pickles10} 
}
\end{table*}
\end{landscape}
    
\setcounter{table}{3}
\begin{landscape}
\textwidth = 700 pt
\begin{table*}
\setlength{\tabcolsep}{.5pt}
{\tiny
\center
\begin{tabular}{clcccccccccccccccccccccc}
\hline
(1) & (2) & (3) & (4) & (5)& (6)& (7)& (8)& (9) & (10)&  (11) & (12) & (13) & (14) & (15) & (16) & (17) & (18) & (19) & (20) & (21) & (22) & (23) & (24)\\
	ID &       Star &          $\alpha$  (J2000) &          $\delta$ (J2000) &          $l$  &          $b$ &    $E_{d}(B-V)$ &          $V$ &       $V_{err}$ &      $(B-V)$ &   $(B-V)_{err}$ &        $V-R_c$ &   $(V-R_c)_{err}$ &        Ref &
         $g'$ &     $g'_{err}$ &      $(g'-r')$ &  $(g'-r')_{err}$ &      $(r'-i')$ &  $(r'-i')_{err}$ &       $T_{eff}$ &       $\log g$  & $[Fe/H]$ & Ref \\

	&       &          (hh:mm:ss.ss) &          (dd:mm:ss.ss) &          (deg) &          (deg)&    (mag) &          (mag)  &       (mag) &      (mag) &   (mag) &     (mag) &   (mag) &        &
         (mag) &     (mag) &      (mag) &  (mag) &      (mag) &  (mag) &       (K) &       (cms$^{-2}$) &(dex) &      \\
\hline
	61 & NGC 6341 12018 & 17 17 22.38 & +43 06 56.34 &    68.320 &     34.811 &      0.019 &     14.432 &      0.008 &      0.672 &      0.013 &      0.428 &      0.011 &          1 &     14.849 &      0.009 &      0.564 &      0.011 &      0.220 &      0.008 &       5160 &       2.30 & -2.34 & 2000AJ....120.1351S \\
        62 &  HD 170737    & 18 29 54.11 & +26 39 26.24 &     55.019 &     16.213 &      0.043 &      8.117 &      0.011 &      0.804 &      0.011 &      0.462 &      0.011 &          1 &      8.472 &      0.001 &      0.597 &      0.001 &      0.240 &      0.001 &       5100 &       3.30 &-0.68 &2012A\&A...541A.157S \\
        63 & BD+05 3839    & 18 37 34.21 & +05 28 33.46 &     36.243 &      5.572 &      0.229 &      9.397 &      0.002 &      1.190 &      0.003 &      0.606 &      0.003 &          1 &      9.904 &      0.002 &      0.906 &      0.003 &      0.371 &      0.003 &       5100 &       2.83 &  -0.01 &1994ApJS...91..309L \\
        64 & BD+05 3858    & 18 38 20.75 & +05 26 02.31 &     36.293 &      5.380 &      0.207 &      9.358 &      0.002 &      1.066 &      0.003 &      0.534 &      0.003 &          1 &      9.804 &      0.002 &      0.788 &      0.003 &      0.329 &      0.003 &       5200 &       3.00 & -0.03 & 1994ApJS...91..309L \\
        65 &  HD 175305    & 18 47 06.44 & +74 43 31.45 &    105.834 &     26.379 &      0.048 &      7.252 &      0.001 &      0.772 &      0.001 &      0.473 &      0.002 &          1 &      7.543 &      0.001 &      0.569 &      0.001 &      0.196 &      0.001 &       5036 &       2.76 & -1.35 & 2012ApJ...753...64I \\
        66 & NGC 6705 1423 & 18 50 55.82 & -06 18 14.80 &     27.259 &     -2.758 &      0.318 &     11.454 &      0.003 &      1.634 &      0.006 &      0.853 &      0.004 &          1 &     12.178 &      0.003 &      1.316 &      0.004 &      0.513 &      0.002 &       4750 &       2.90 &0.04 &2006AJ....131.2949S \\
        67 & NGC 6705 1256 & 18 51 00.24 & -06 16 59.50 &     27.286 &     -2.764 &      0.372 &     11.626 &      0.004 &      1.699 &      0.009 &      0.909 &      0.004 &          1 &     12.387 &      0.004 &      1.403 &      0.004 &      0.572 &      0.003 &       4600 &       2.50 & 0.28 &2006AJ....131.2949S \\
        68 & NGC 6705 1223 & 18 51 00.93 & -06 14 56.40 &     27.318 &     -2.751 &      0.356 &     11.480 &      0.003 &      1.230 &      0.005 &      0.693 &      0.004 &          1 &     12.030 &      0.003 &      1.006 &      0.004 &      0.423 &      0.003 &       4750 &       2.50 & -0.06 &2006AJ....131.2949S \\
        69 & BD+05 4314    & 19 51 49.60 & +05 36 45.84 &     44.989 &    -10.725 &      0.119 &     10.757 &      0.002 &      0.779 &      0.003 &      0.459 &      0.003 &          1 &     11.079 &      0.002 &      0.575 &      0.003 &      0.218 &      0.003 &       5000 &       3.00 & -1.51 &2003PASP..115...22Y \\
        70 & NGC 6838 1056 & 19 53 48.40 & +18 48 23.50 &     56.773 &     -4.557 &      0.257 &     13.251 &      0.005 &      1.339 &      0.010 &      0.741 &      0.006 &          1 &     13.931 &      0.004 &      1.124 &      0.004 &      0.485 &      0.003 &       4582 &       2.10 &-0.77 &1986A\&A...169..208G \\
        71 & BD-14 5890    & 20 56 09.13 & -13 31 17.66 &     34.408 &    -33.683 &      0.030 &     10.192 &      0.002 &      0.803 &      0.003 &      0.478 &      0.002 &          1 &     10.517 &      0.001 &      0.653 &      0.002 &      0.230 &      0.003 &       4891 &       2.03 &-2.16 &2012ApJ...753...64I \\
        72 & BD-03 5215    & 21 28 01.31 & -03 07 40.93 &     50.075 &    -35.866 &      0.048 &     10.123 &      0.005 &      0.613 &      0.005 &      0.368 &      0.006 &          1 &     10.384 &      0.004 &      0.427 &      0.005 &      0.176 &      0.005 &       5478 &       2.17 & -1.49 &2012ApJ...753...64I \\
        73 & BPS CS 22944-032 & 21 47 43.10 & -13 40 22.00 &  40.906 &    -45.175 &      0.038 &     13.161 &      0.003 &      0.597 &      0.004 &      0.390 &      0.004 &          1 &     13.433 &      0.002 &      0.482 &      0.003 &      0.196 &      0.003 &       5300 &       2.87 &-2.98 &2008ApJ...681.1524L \\
        74 &  HD 210295    & 22 09 41.44 & -13 36 19.47 &     44.406 &    -49.964 &      0.035 &      9.455 &      0.002 &      0.847 &      0.002 &      0.469 &      0.002 &          1 &      9.937 &      0.001 &      0.719 &      0.001 &      0.254 &      0.002 &       4763 &       2.19 &-1.25 &2012ApJ...753...64I \\
        75 &  HD 219715    & 23 18 01.19 & +09 04 28.11 &     87.675 &    -47.301 &      0.050 &      9.197 &      0.002 &      0.779 &      0.003 &      0.456 &      0.003 &          1 &      9.733 &      0.004 &      0.566 &      0.005 &      0.213 &      0.004 &       5000 &       2.50 &-1.10 &2000A\&A...353..978M \\
        76 & NGC 7789 329  & 23 56 55.46 & +56 45 09.10 &    115.477 &     -5.328 &      0.313 &     12.307 &      0.003 &      1.433 &      0.007 &      0.774 &      0.004 &          1 &     12.999 &      0.005 &      1.208 &      0.005 &      0.489 &      0.003 &       4345 &       2.20 & 0.10 &1985PASP...97..801P \\
        77 & NGC 7789 353  & 23 56 57.52 & +56 45 27.30 &    115.483 &     -5.324 &      0.326 &     12.612 &      0.004 &      1.432 &      0.009 &      0.784 &      0.004 &          1 &     13.303 &      0.006 &      1.206 &      0.007 &      0.487 &      0.004 &       4345 &       2.20 & 0.15 &1985PASP...97..801P \\
        78 & NGC 7789 637  & 23 57 22.43 & +56 41 46.00 &    115.526 &     -5.396 &      0.313 &     12.419 &      0.003 &      1.457 &      0.008 &      0.784 &      0.004 &          1 &     13.118 &      0.006 &      1.222 &      0.007 &      0.513 &      0.004 &       4383 &       2.10 &0.00 &1985PASP...97..801P \\
        79 & NGC 7789 737  & 23 57 29.97 & +56 43 19.90 &    115.548 &     -5.374 &      0.327 &     13.383 &      0.007 &      1.179 &      0.016 &      0.645 &      0.008 &          1 &     13.996 &      0.013 &      0.999 &      0.014 &      0.396 &      0.008 &       4990 &       2.80 & -0.10 &1985PASP...97..801P \\
        80 & NGC 7789 765  & 23 57 31.87 & +56 41 22.12 &    115.546 &     -5.407 &      0.275 &     11.634 &      0.002 &      1.499 &      0.004 &      0.800 &      0.002&          1 &     12.325 &      0.003 &      1.223 &      0.003 &      0.515 &      0.001 &       4383 &       2.10 &-0.20 &1985PASP...97..801P \\
\hline
\end{tabular}
(1) This study, (2) \citet{Pickles10} 
}
\end{table*}
\end{landscape}

\subsection{Distances and De-reddening of the Magnitudes and Colours}

After obtaining standardized observed magnitudes and colours of the present sample of giants (80) through Eqs. 1 to 6, de-reddening of them is the next step before attempting to establish the transformation equations between $BVR_c$ and $g'r'i'$. A priory advantage for us to know calibrated absolute magnitudes of giants from the tables of \citet{Sung13}. \citet{Sung13} have studied spectral type-$M_V$ and spectral type-$T_{eff}$ relation of giants in general along with other luminosity classes on the H-R diagram in the Johnson-Cousins $UBVR_cI_c$ system. Using their Table 4 and 5, we have plotted absolute magnitudes as a function of the effective temperature. The data existing on these tables in the range of $3600<T_{eff}(K)\leq16000$ were shown by filled circles in Fig. 1. In order to increase efficient use of the data, two polynomial functions were fitted to the cooler ($3600<T_{eff}(K)\leq6000$) and to the hotter ($6000<T_{eff} (K)<16000$) regions. Dashed line in Fig. 1 represents the polynomials fitted. 

For a given effective temperature of our sample giants as supplied by the Pastel catalogue \citep{Soubiran10}, the absolute magnitude ($M_V$) is provided by Fig. 1. Nevertheless, absolute  magnitude alone is not enough to estimate neither the distance nor the interstellar extinction. The total interstellar absorption in $V$-band could be estimated by means of the maps of \citet{Schlafly11} which are based on a recalibration of the \citet*{Schlegel98} maps, for each of the sample star, by including the Galactic coordinates of the star in question into the NED service\footnote{http://ned.ipac.caltech.edu/forms/calculator.html}. The value $A_{\infty}$ is valid for a star at infinite distance. However, we used it in the Pogson's equation and evaluated the distance to the star as a first approximation:

\begin{equation}
V-M_V-A_{\infty}=5\log d-5,
\end{equation}
where $V$ and $M_V$ are the apparent and the absolute magnitudes from which the distance is estimated. We then reduced $A_{\infty}$ to the total absorption of the star at distance $d$, i.e. $A_d$, by the procedure of \citet{Bahcall80}. We have replaced the numerical value of $A_d(b)$ with the one of $A_{\infty}$ in Eq. (7), and applied a series of iteration to obtain the final total absorption $A_V$ by which the selective absorption (colour excess), $E_d(B-V)$ could be evaluated for distance $d$ as follows:

\begin{equation}
E_{d}(B-V)=A_{d}(b)~/~3.1.
\end{equation}

$E_d(B-V)$ is the standard $E(B-V)$ colour excess of the star with given $V$, $M_V$ and $d$, from which one can compute intrinsic colour by using observed colour and de-reddened observed magnitude. The range of the colour excess of the sample stars is $0 < E(B-V) < 0.84$ mag, and their distribution in the 
Galactic longitude-Galactic latitude plane is plotted in Fig. 2, where one may notice the $E(B-V)$ colour excess is highest on the Galactic plane and decreases towards the Galactic poles. We have made consistency check of $E(B-V)$ colour excesses of 25 giant stars by comparing the colour excesses of the clusters which they belong to. The results are given in Table 4, where the columns are explanatory to indicate the name of the cluster, the star's ID, equatorial coordinates (J2000), the colour excesses $E_{d}(B-V)$ of the stars estimated in this study and the colour excess $E_{cl}(B-V)$ of the cluster from the literature. The colour excesses $E_{cl}(B-V)$ of the clusters NGC 6341 and NGC 6838 are taken from \citet[][edition 2010]{Harris96}, while those for the remaining five clusters are provided from \citet{Dias02}. Table 4 shows that the colour excesses evaluated in this study and taken from literature are in good agreement, in general.   

The magnitudes and colours are de-reddened by using the corresponding $E(B-V)$ colour excess of the star and the equations in the literature, i.e. we adopted $A_V$/$E(B-V)$=3.1, $E(V-R_c)/E(B-V)$=0.65 \citep*{Cardelli89} and $A_{g'}/A_V$=1.199, $A_{r'}/A_V$=0.858, $A_{i'}/A_V$=0.639 \citep{Fan99}. The two colour-diagrams for the sample stars are plotted with symbols in Fig. 3 and Fig. 4 for Johnson-Cousins and SDSS systems, respectively. The solid curves in Fig. 3 and Fig. 4 are adopted from the synthetic data in \citet{Pickles98} and in \citet{Covey07}, respectively. The mean errors of the observed magnitude and colours of both systems are given in Table 5 and plotted in Fig. 5. 

\begin{table*}
\setlength{\tabcolsep}{4pt}
\center
\caption{The colour excesses of the stars observed in various stellar clusters. Cluster names, star IDs, equatorial coordinates and evaluated colour excesses ($E_d(B-V)$) of stars are given in columns 1-5. The last two columns include colour excesses ($E_{cl}(B-V)$) of the clusters and their reference.}
\begin{tabular}{ccccccc}
\hline
   Cluster &       Star &  $\alpha$ (J2000) &  $\delta$ (J2000) & $E_d(B-V)$ & $E_{cl}(B-V)$ &  Reference \\
           &            & (hh:mm:ss.ss) & (dd:mm:ss.ss) &      (mag) &      (mag) &            \\
\hline
  NGC 2112 &        204 & 05 53 47.54 & +00 22 02.00 &      0.793 &      0.600 & \citet{Dias02} \\
           &        402 & 05 53 53.30 & +00 24 37.90 &      0.839 &      0.600 & \citet{Dias02} \\
  NGC 2420 &        115 & 07 38 21.67 & +21 33 51.40 &      0.036 &      0.040 & \citet{Dias02} \\
           &        173 & 07 38 26.96 & +21 33 31.30 &      0.036 &      0.040 & \citet{Dias02} \\
  NGC 2682 &         84 & 08 51 12.70 & +11 52 42.40 &      0.030 &      0.040 & \citet{Dias02} \\
           &        105 & 08 51 17.10 & +11 48 16.10 &      0.030 &      0.040 & \citet{Dias02} \\
           &        141 & 08 51 22.80 & +11 48 01.70 &      0.030 &      0.040 & \citet{Dias02} \\
           &        151 & 08 51 26.19 & +11 53 52.00 &      0.031 &      0.040 & \citet{Dias02} \\
           &        164 & 08 51 28.99 & +11 50 33.10 &      0.031 &      0.040 & \citet{Dias02} \\
           &        224 & 08 51 43.55 & +11 44 26.40 &      0.025 &      0.040 & \citet{Dias02} \\
           &        231 & 08 51 45.08 & +11 47 45.90 &      0.027 &      0.040 & \citet{Dias02} \\
           &        266 & 08 51 59.52 & +11 55 04.90 &      0.032 &      0.040 & \citet{Dias02} \\
  NGC 6341 &       9012 & 17 16 52.84 & +43 03 29.50 &      0.020 &      0.020 & \citet{Harris96} \\
           &      10065 & 17 17 11.41 & +43 06 02.70 &      0.019 &      0.020 & \citet{Harris96} \\
           &      11027 & 17 17 21.61 & +43 06 15.90 &      0.019 &      0.020 & \citet{Harris96} \\
           &      12018 & 17 17 22.38 & +43 06 56.34 &      0.019 &      0.020 & \citet{Harris96} \\
  NGC 6705 &       1423 & 18 50 55.82 &$-$06 18 14.80&      0.318 &      0.428 & \citet{Dias02} \\
           &       1256 & 18 51 00.24 &$-$06 16 59.50&      0.372 &      0.428 & \citet{Dias02} \\
           &       1223 & 18 51 00.93 &$-$06 14 56.40&      0.356 &      0.428 & \citet{Dias02} \\
  NGC 6838 &       1056 & 19 53 48.40 & +18 48 23.50 &      0.257 &      0.250 & \citet{Harris96} \\
  NGC 7789 &        329 & 23 56 55.46 & +56 45 09.10 &      0.313 &      0.280 & \citet{Dias02} \\
           &        353 & 23 56 57.52 & +56 45 27.30 &      0.326 &      0.280 & \citet{Dias02} \\
           &        637 & 23 57 22.43 & +56 41 46.00 &      0.313 &      0.280 & \citet{Dias02} \\
           &        737 & 23 57 29.97 & +56 43 19.90 &      0.327 &      0.280 & \citet{Dias02} \\
           &        765 & 23 57 31.87 & +56 41 22.12 &      0.275 &      0.280 & \citet{Dias02} \\
\hline
\end{tabular}
\end{table*}

\begin{table}
\setlength{\tabcolsep}{4pt}
\center
\caption{Mean values of the errors for magnitude and colours in 
the $BVR_c$ and $g'r'i'$ systems.}
\begin{tabular}{lclc}
\hline
Magnitude/   & Mean Error & Magnitude/ & Mean Error \\
Colour       & (mag) & Colour     & (mag)\\
\hline
$V$	& 0.003 & $g'$		& 0.003 \\
$B-V$	& 0.005 & $g'-r'$	& 0.003 \\
$V-R_c$	& 0.004 & $r'-i'$	& 0.003 \\
\hline
\end{tabular}
\end{table}

\section{Transformations}
The following general equations have been preferred to derive the 18 sets of transformation equations with their coefficients between $BVR_c$ and $g'r'i'$ photometric systems using de-reddened magnitudes and colours of the sample giants in this study. Eqs. (9)-(17) transform $BVR_c$ into $g'r'i'$ magnitudes and colours, while Eqs. (18)-(26) are their inverse transformations. The
equations are: 

\begin{eqnarray}
(g'-V)_{0}=a_{i}(B-V)_{0}^2+b_{i}(B-V)_{0}+c_{i}\\
(g'-V)_{0}=a_{i}(V-R_c)_{0}^2+b_{i}(V-R_c)_{0}+c_{i}\\
(g'-V)_{0}=a_{i}(B-V)_{0}+b_{i}(V-R_c)_{0}+c_{i}	\\
(g'-r')_{0}=a_{i}(B-V)_{0}^2+b_{i}(B-V)_{0}+c_{i}\\
(g'-r')_{0}=a_{i}(V-R_c)_{0}^2+b_{i}(V-R_c)_{0}+c_{i}\\
(g'-r')_{0}=a_{i}(B-V)_{0}+b_{i}(V-R_c)_{0}+c_{i}	\\
(r'-i')_{0}=a_{i}(B-V)_{0}^2+b_{i}(B-V)_{0}+c_{i}\\
(r'-i')_{0}=a_{i}(V-R_c)_{0}^2+b_{i}(V-R_c)_{0}+c_{i}\\
(r'-i')_{0}=a_{i}(B-V)_{0}+b_{i}(V-R_c)_{0}+c_{i}\\
\nonumber
\end{eqnarray}

\begin{eqnarray}
(V-g')_{0}=d_{i}(g'-r')_{0}^2+e_{i}(g'-r')_{0}+f_{i}\\
(V-g')_{0}=d_{i}(r'-i')_{0}^2+e_{i}(r'-i')_{0}+f_{i}\\
(V-g')_{0}=d_{i}(g'-r')_{0}+e_{i}(r'-i')_{0}+f_{i}\\
(B-V)_{0}=d_{i}(g'-r')_{0}^2+e_{i}(g'-r')_{0}+f_{i}\\
(B-V)_{0}=d_{i}(r'-i')_{0}^2+e_{i}(r'-i')_{0}+f_{i}\\
(B-V)_{0}=d_{i}(g'-r')_{0}+e_{i}(r'-i')_{0}+f_{i}\\
(V-R_c)_{0}=d_{i}(g'-r')_{0}^2+e_{i}(g'-r')_{0}+f_{i}\\
(V-R_c)_{0}=d_{i}(r'-i')_{0}^2+e_{i}(r'-i')_{0}+f_{i}\\
(V-R_c)_{0}=d_{i}(g'-r')_{0}+e_{i}(r'-i')_{0}+f_{i}.\\
\nonumber
\end{eqnarray}
The values of the coefficients and their errors are given in Table 6 and Table 7. The distribution of the sample stars on the colour planes are plotted in Fig. 6. The ranges of the magnitudes and colours in the transformations are: $7.10<V_0<14.50$, $7.30<g_0<14.85$, $-0.20<(B-V)_0<1.41$, $-0.11<(V-R_c)_0<0.73$, $-0.42<(g-r)_0<1.15$, and $-0.37<(r-i)_0<0.47$ mag. The residuals of the colours from the curve fits are plotted in Fig. 7 and Fig. 8, while the means and standard deviations of the residuals are included to the data in Tables 6 and 7.

The transformation equations from $BVR_c$ to $g'r'i'$ colours derived in this study have been applied to another sample of 427 giants taken from \citet{Pickles10} as explained in the following. \citet{Smith05} have published a list of $\sim$16000 southern SDSS standards, which includes some repetition. 6117 of them were turned out to be of luminosity class III, i.e. giants, which cover the spectral range A-M. We have selected a sample of 427 giants with uncertainties in $g'$, $r'$ and $i'$ less than 0.01 mag. The $E(B-V)$ colour excesses used for de-reddening the $g'$ magnitudes, and $g'-r'$ and $r'-i'$ colours have been evaluated in two steps, by a procedure similar to the one used for the original sample from which the Eqs. 7 and 8 derived as in Section 2.2. The $g'$ magnitudes, and $g'-r'$ and $r'-i'$ colours are de-reddened according to the corresponding $E(B-V)$ colour excesses of the stars and the equations of \cite{Fan99}. The $(g'-r')_0 \times (r'-i')_0$ two-colour diagram of the sample of 427 stars obtained from \citet{Pickles10} are given in Fig. 9.

To estimate intrinsic $BVR_c$ colours of the sample of those 427 giants, we have identified them first in \citet{Pickles10}, and then recorded their $V$ magnitudes and, $B-V$ and $V-R_c$ colours. Then, those $V$ magnitudes and, $B-V$ and $V-R_c$ colours are de-reddened by the procedure explained in the Section 2.2, and transformed them to $g'_0$ magnitudes and, $(g'-r')_0$ and $(r'-i')_0$ colours by using the transformation equations derived from the original sample of 80 giants in this study. Then, we have compared those $g'_0$ magnitude and, $(g'-r')_0$ and $(r'-i')_0$ colours obtained through transformation equations to the $g'_0$ magnitude and, $(g'-r')_0$ and $(r'-i')_0$ colours evaluated from \citet{Pickles10}. The residuals, i.e. the differences between the original and evaluated colours, are plotted in Fig. 10. The mean and standard deviations of the residuals for each colour are also indicated in the corresponding panel of the figure. 

\begin{table*}
\setlength{\tabcolsep}{3pt}
\center
\caption{Coefficients for the Eqs. (9)-(17). The figures in the first line indicate the equation number, 
$R$ is the correlation coefficient and, $s$ and $m.r.$ are the standard deviation, and mean residuals, 
respectively.}   
\begin{tabular}{cccccc}
\hline
     &   (9) &    (10) &   (11) &   (12) &   (13) \\
\hline
Coefficient & $(g'-V)_0$ & $(g'-V)_0$ & $(g'-V)_0$ & $(g'-r')_0$ & $(g'-r')_0$ \\
\hline
        $a_i$ & -0.122$\pm$0.051 & -0.155$\pm$0.227 & 0.404$\pm$0.083 & -0.045$\pm$0.031 & 0.381$\pm$0.120 \\
        $b_i$ &  0.651$\pm$0.074 &  1.044$\pm$0.160 & 0.168$\pm$0.166 & 1.006$\pm$0.055 & 1.528$\pm$0.055 \\
        $c_i$ & -0.069$\pm$0.028 & -0.056$\pm$0.032 & -0.035$\pm$0.024 & -0.197$\pm$0.024 & -0.170$\pm$0.009 \\
          $R$ &      0.912 &      0.876 &      0.907 &      0.986 &      0.992 \\
          $s$ &      0.068 &      0.080 &      0.070 &      0.059 &      0.059 \\
       $m.r.$ &      0.000 &      0.000 &      0.000 &     -0.006 &     -0.002 \\
\hline
    &   (14) &   (15) &   (16) &   (17) &    \\
\hline
Coefficient & $(g'-r')_0$ & $(r'-i')_0$ & $(r'-i')_0$ & $(r'-i')_0$ &            \\
\hline
        $a_i$ &  0.464$\pm$0.063 & -0.243$\pm$0.035 & -0.420$\pm$0.064 & -0.026$\pm$0.049 &  \\
        $b_i$ &  0.931$\pm$0.128 &  0.730$\pm$0.051 &  1.111$\pm$0.030 &  0.897$\pm$0.099 &  \\
        $c_i$ & -0.205$\pm$0.018 & -0.224$\pm$0.019 & -0.209$\pm$0.005 & -0.185$\pm$0.014 &  \\
          $R$ &      0.982 &      0.940 &      0.993 &      0.953 &            \\
          $s$ &      0.054 &      0.047 &      0.031 &      0.042 &            \\
       $m.r.$ &      0.000 &      0.000 &     -0.001 &      0.000 &            \\
\hline
\end{tabular}  
\end{table*}

\begin{table*}
\setlength{\tabcolsep}{3pt}
\center
\caption{Coefficients for the Eqs. (18)-(26). The symbols are as in Table 6.}   
\begin{tabular}{cccccc}
\hline
		&       (18)	&       (19) &        (20)&        (21) &        (22) \\
\hline
Coefficient 	& $(V-g')_0$ & $(V-g')_0$ & $(V-g')_0$ & $(B-V)_0$ & $(B-V)_0$ \\
\hline
	$d_i$	&  0.041$\pm$0.016 	& -0.104$\pm$0.165 & -0.460$\pm$0.059 & 0.022$\pm$0.018 & 0.466$\pm$0.117 \\
        $e_i$	& -0.615$\pm$0.013 	& -1.096$\pm$0.029 & -0.243$\pm$0.111 & 1.028$\pm$0.014 & 1.947$\pm$0.028 \\
        $f_i$	& -0.030$\pm$0.007 	& -0.156$\pm$0.014 & -0.052$\pm$0.015 & 0.205$\pm$0.007 & 0.418$\pm$0.015 \\
         $R$ 	&      0.989 &      0.977 &      0.989 &      0.996 &      0.987 \\
         $s$ 	&      0.049 &      0.067 &      0.050 &      0.055 &      0.098 \\
      $m.r.$ 	&      0.001 &     -0.001 &     -0.005 &      0.002 &      0.005 \\
\hline
		&       (23) &       (24) &       (25) &        (26) &            \\
\hline
Coefficient 	&     $(B-V)_0$    &     $(V-R_c)_0$  &     $(V-R_c)_0$ &     $(V-R_c)_0$ &  \\
\hline
        $d_i$ 	&  1.321$\pm$0.092 & -0.002$\pm$0.016 & 0.102$\pm$0.097 & 0.335$\pm$0.037 &  \\
        $e_i$ 	& -0.616$\pm$0.193 &  0.552$\pm$0.009 & 1.085$\pm$0.017 & 0.449$\pm$0.068 &  \\
        $f_i$ 	&  0.162$\pm$0.023 &  0.112$\pm$0.004 & 0.229$\pm$0.009 & 0.158$\pm$0.009 &  \\
         $R$ 	&      0.974 &      0.992 &      0.991 &      0.995 &            \\
         $s$ 	&      0.071 &      0.033 &      0.039 &      0.030 &            \\
      $m.r.$ 	&      0.000 &      0.007 &      0.001 &      0.002 &            \\
\hline
\end{tabular}  
\end{table*}

\section{Summary and Discussion}

The transformation equations from $BVR_c$ to $g'r'i'$ colours and vice versa for giants were derived and presented. The transformation equations were obtained using the observed magnitudes and colours of a sample of 80 giants selected from the Pastel catalogue \citep{Soubiran10} with confirmed surface gravity ($2\leq \log g$ (cms$^{-2}$) $\leq3$) at effective temperatures from 4000 to 16000 K. The $g'r'i'$ magnitudes of all sample and 65 of them at $BVR_c$ magnitudes were obtained by observations carried out with the T100 telescope at TUG at Bak{\i}rl{\i}tepe, Antalya, in the years 2011-2012. The $BVR_c$ magnitudes of 15 giants were completed from the synthetic $BVR_c$ magnitudes in \citet{Pickles10}. We have used the $M_V$ absolute magnitudes and $T_{eff}$ temperatures for giants at various spectral types presented in \citet{Sung13} to produce Fig. 1, which were used to estimated the $M_V$ absolute magnitudes of the sample stars used in this study. We have de-reddened the magnitude and colours in both photometric systems, $BVR_c$ and $g'r'i'$ of the sample by a procedure commonly used in the literature.

The transformation equations from $BVR_c$ to $g'r'i'$ (Eqs. 9-17) and inverse transformations (Eqs. 18-26) were derived by fitting quadratic polynomials to $(g'-V)_0$, $(g'-r')_0$ and $(r'-i')_0$ which are given in terms of $(B-V)_0$ only or  $(V-R_c)_0$ only, or by fitting a linear function to them if they are expressed both $(B-V)_0$ and $(V-R_c)_0$. Similarly, $(V-g')_0$, $(B-V)_0$ and $(V-R_c)_0$ expressed by a quadratic function of $(g'-r')_0$ or $(r'-i')_0$, or they are expressed by a linear function fit both $(g'-r')_0$ and $(r'-i')_0$ colours used in the equation. The correlation coefficients and the standard deviations in Table 6 show that all the transformation equations (Eqs. 9-17) provide sufficiently accurate magnitude and colours. However, the most accurate colours $(g'-V)_0$, $(g'-r')_0$ and $(r'-i')_0$ come from the quadratic equations containing only $(B-V)_0$, or $(V-R_c)_0$. The correlation coefficients given in Table 7 for inverse transformations are a bit larger, while the standard deviations are smaller. The most accurate colours $(V -g')_0$, $(B-V)_0$, and $(V-R)_0$ come from the quadratic equations containing only $(g'-r')_0$, or $(g'-r')_0$ and from the linear equation if both $(g'-r')_0$ and $(g'-r')_0$ appear in the equation.

The transformation equations could be considered valid for the ranges of the magnitudes and colours used in the transformations: $7.10<V_0<14.50$, $7.30<g'_0<14.85$,$-0.20<(B-V)_0<1.41$, $-0.11<(V-R_c)_0<0.73$, $-0.42<(g'-r')_0<1.15$, and $-0.37<(r'-i')_0<0.47$ mag, rather larger than the ones provided by \citet{Yaz10}, i.e. $0.25<(B-V)_0<1.35$, $0.10<(g-r)_0<0.95$, and $0<(r-i)_0<0.35$. 

We applied the transformation equations derived in this study to the synthetic $BVR_c$ data of 427 giants taken from \citet{Pickles10}. The ranges of the mean residuals ($m.r.$) and standard deviations ($s$) are $-0.010\leq m.r.\leq 0.042$ and $0.028\leq s \leq0.068$ mag, respectively. That is, the $g'$ magnitude and, $(g'-r')_0$ and $(r'-i')_0$ colours of a giant can be estimated by our transformations with an accuracy of at least $\sim$0.05 mag. \citet{Yaz10} derived transformations between 2MASS, SDSS and $BVI$ photometric systems for late type giants for two cases, i.e. metallicity dependent and free of metallicity. We have compared their residuals and the standard deviations free of metallicity. Their mean residuals are rather small, i.e. $-0.0005 \leq m.r. \leq 0.0004$ mag. However, the corresponding standard deviations are larger than the ones in our study, $0.075\leq s\leq 0.167$ mag. That is, our study provides magnitudes and colours with an accuracy of $\sim$1.5 times that of \citet{Yaz10}. 

Sufficiently small residuals and the standard deviations confirm the quality of the observations made in TUG, and accuracy and the precision of the reductions applied to these observations. The metallicities of the sample stars used for deriving the coefficents of the transformation equations between the two photometric systems cover a large range, $-4\leq [Fe/H]\leq 0.5$ dex. However, the number of stars with $[Fe/H]<-1$ dex are small. Hence, we did not consider the metallicity in our transformations. Though, we obtained transformation equations which provide accurate colour and magnitudes.

\section{Acknowledgments}

This work has been supported in part by the Scientific and
Technological Research Council (T\"UB\.ITAK) 212T214.

We thank to T\"UB\.ITAK for a partial support in using T100 telescope with 
project number 11BT100-184-2. 

This research has made use of the NASA/IPAC Infrared Science Archive and 
Extragalactic Database (NED) which are operated by the Jet Propulsion Laboratory, 
California Institute of Technology, under contract with the National Aeronautics 
and Space Administration.

This research has made use of the SIMBAD, and NASA\rq s Astrophysics 
Data System Bibliographic Services.

\begin{figure*}
\begin{center}
\includegraphics[scale=0.60, angle=0]{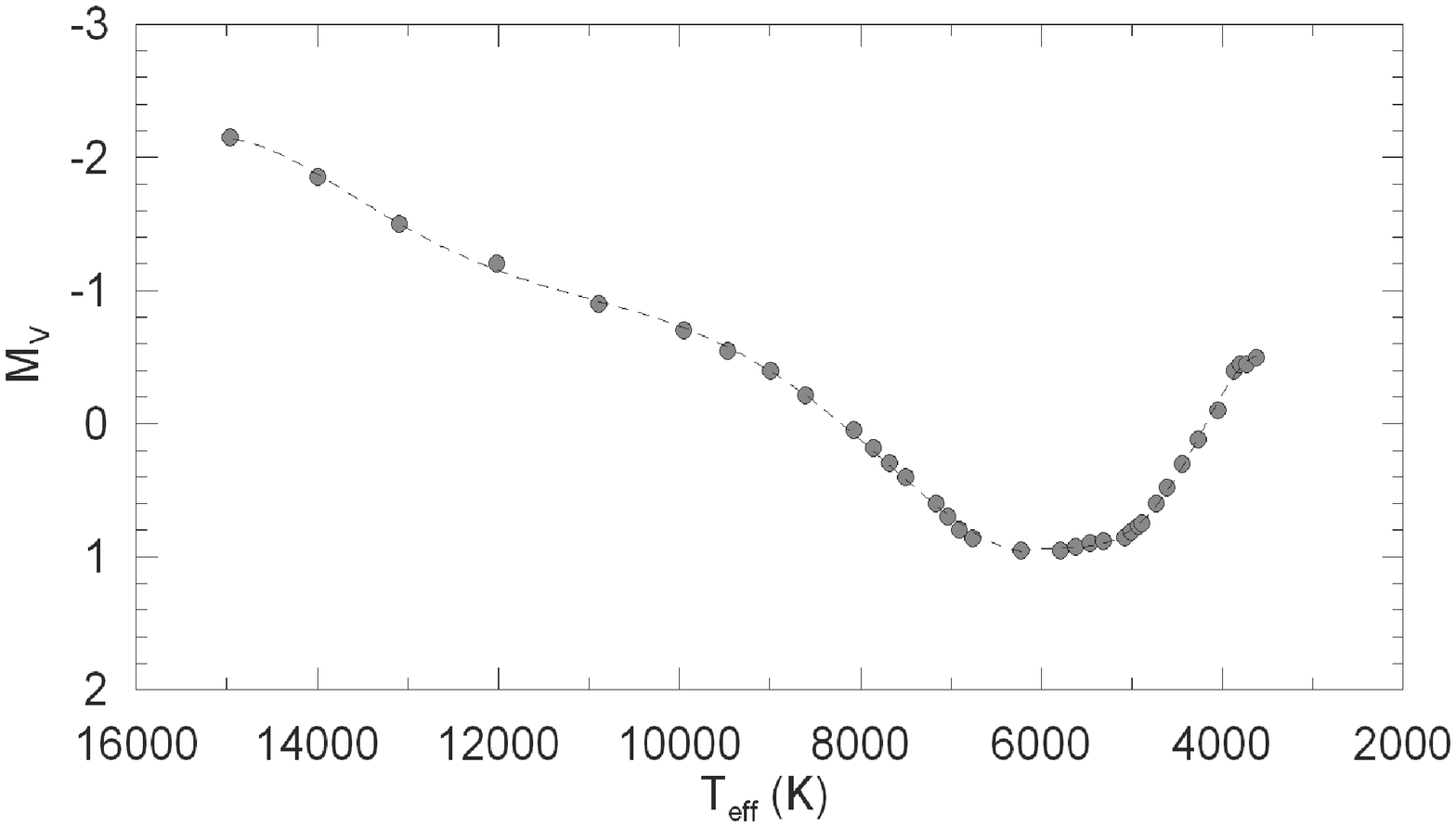}
\caption[] {$M_V \times T_{eff}$ absolute magnitude-temperature diagram for 
the giant stars in \citet{Sung13} which is used for absolute magnitude estimation 
of the sample stars.}
\end{center}
\end{figure*}

\begin{figure*}
\begin{center}
\includegraphics[scale=0.60, angle=0]{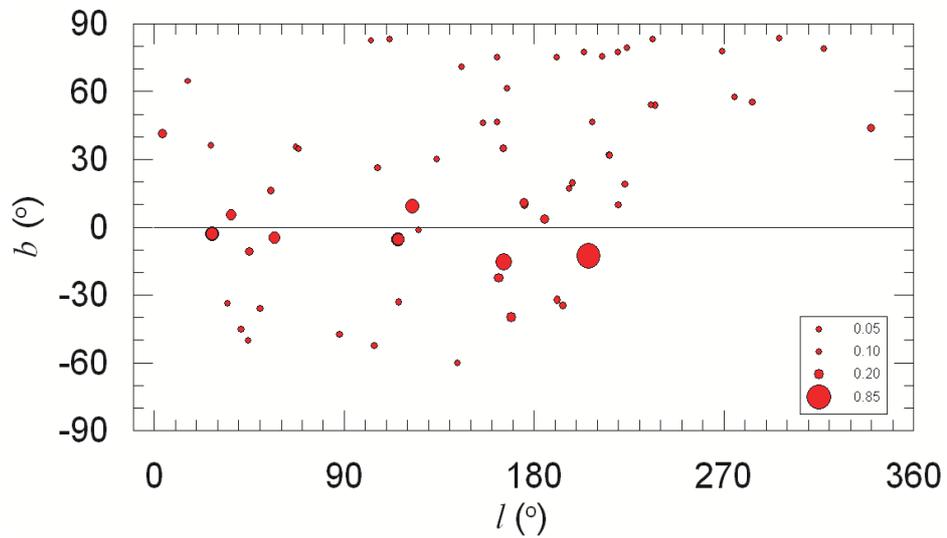}
\caption[] {Galactic coordinates of the programme stars observed in TUG. 
The radius of the circles are proportional to the $E(B-V)$ colour excess 
of the star.}
\end{center}
\end{figure*}

\begin{figure*}
\begin{center}
\includegraphics[scale=0.50, angle=0]{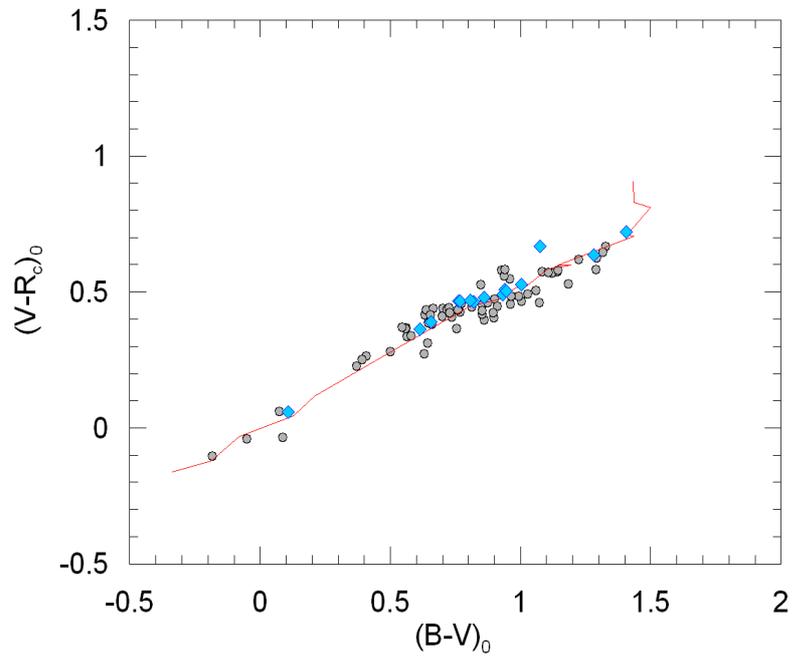}
\caption[] {$(B-V)_0\times(V-R_c)_0$ two-colour diagram of the sample 
stars. The positions of 15 stars with synthetic $(B-V)_0$ and $(V-R_c)_0$ 
colours are marked with a different symbol ($\diamond$). The solid line indicates the 
synthetic two-colour diagram of \citet{Pickles98}.}
\end{center}
\end{figure*}

\begin{figure*}
\begin{center}
\includegraphics[scale=0.50, angle=0]{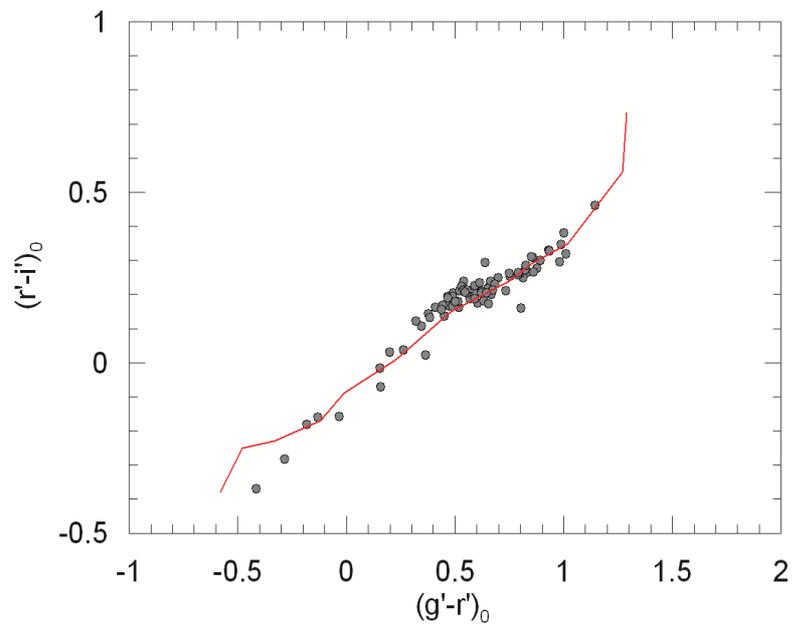}
\caption[] {$(g'-r')_0 \times (r'-i')_0$ two-colour diagram of the 
sample stars. The solid line indicates the synthetic two-colour diagram 
of \citet{Covey07}.}
\end{center}
\end{figure*}

\begin{figure*}
\begin{center}
\includegraphics[scale=0.60, angle=0]{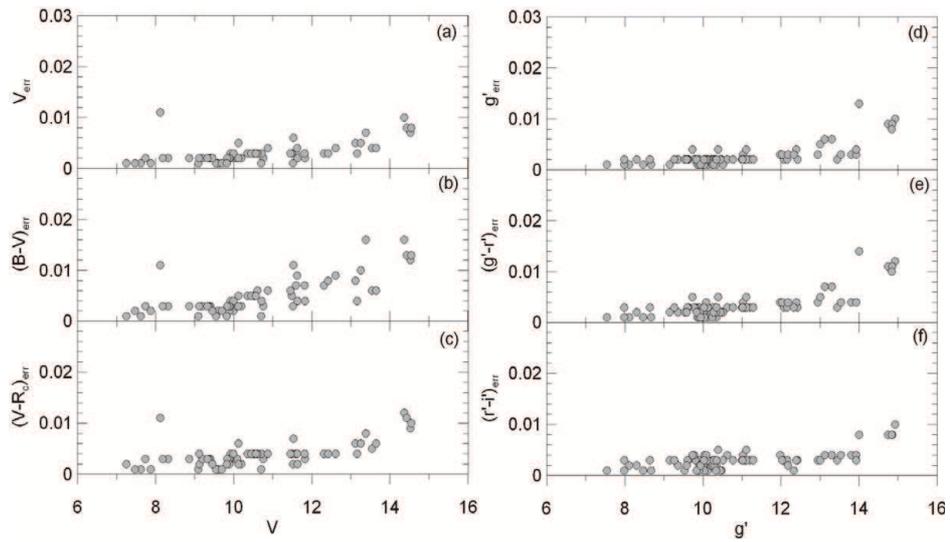}
\caption[] {Distributions of the errors of the magnitudes and colours 
of the sample stars.}
\end{center}
\end{figure*}

\begin{figure*}
\begin{center}
\includegraphics[scale=0.60, angle=0]{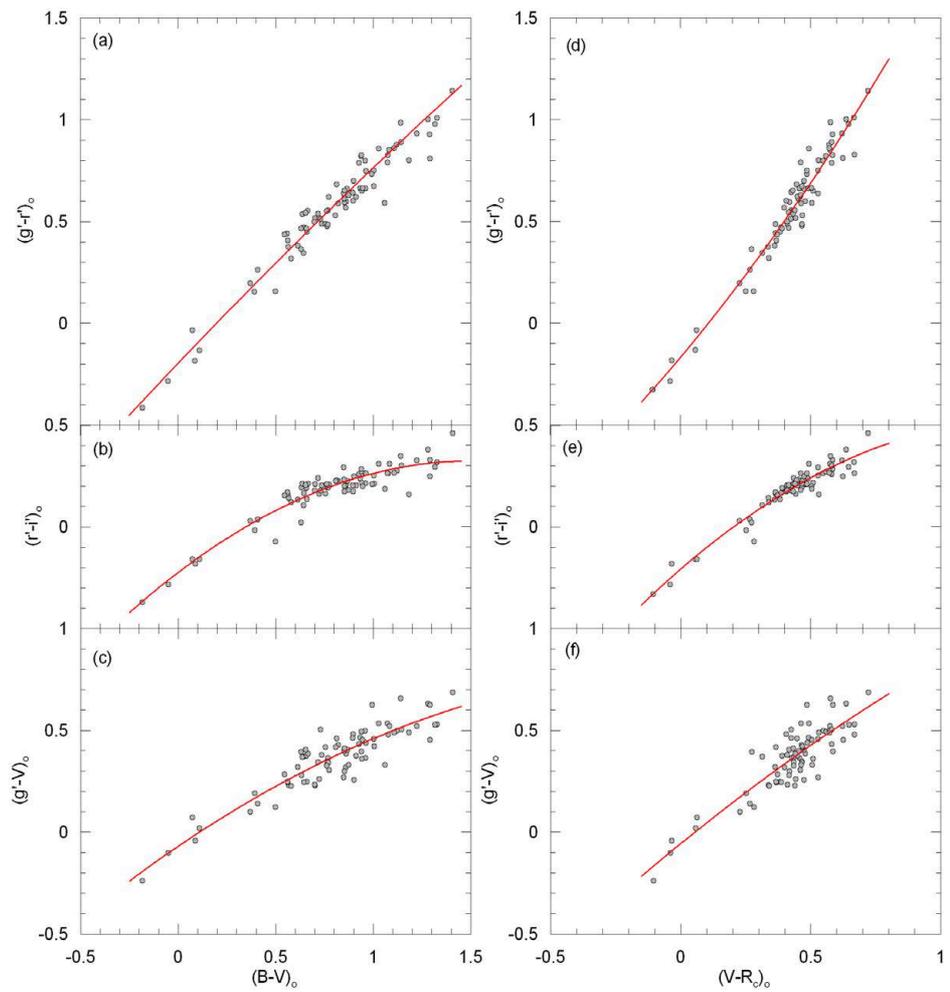}
\caption[] {Distributions of the sample stars in six colour planes. 
The curves indicate quadratic polynomials.}
\end{center}
\end{figure*}

\begin{figure*}
\begin{center}
\includegraphics[scale=0.85, angle=0]{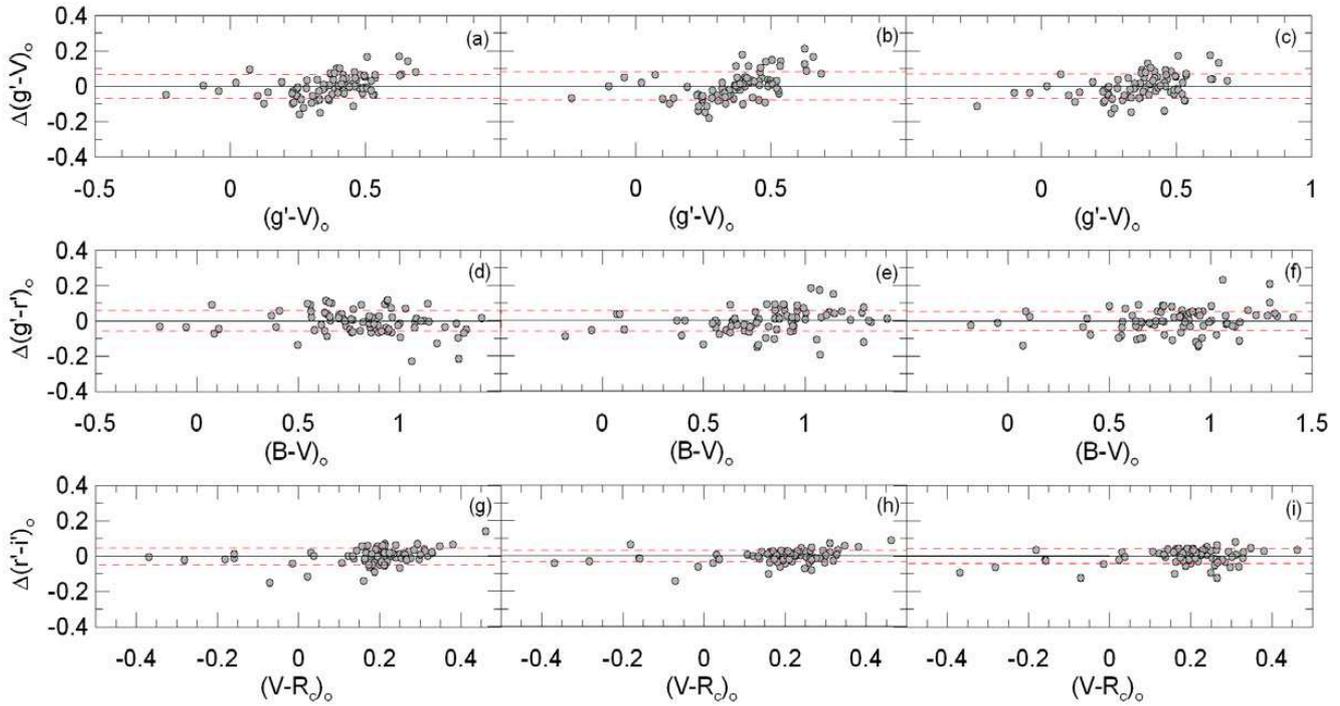}
\caption[] {Distributions of the residuals of the sample stars for 
transformation equations (Eqs. 9-17).}
\end{center}
\end{figure*}

\begin{figure*}
\begin{center}
\includegraphics[scale=0.85, angle=0]{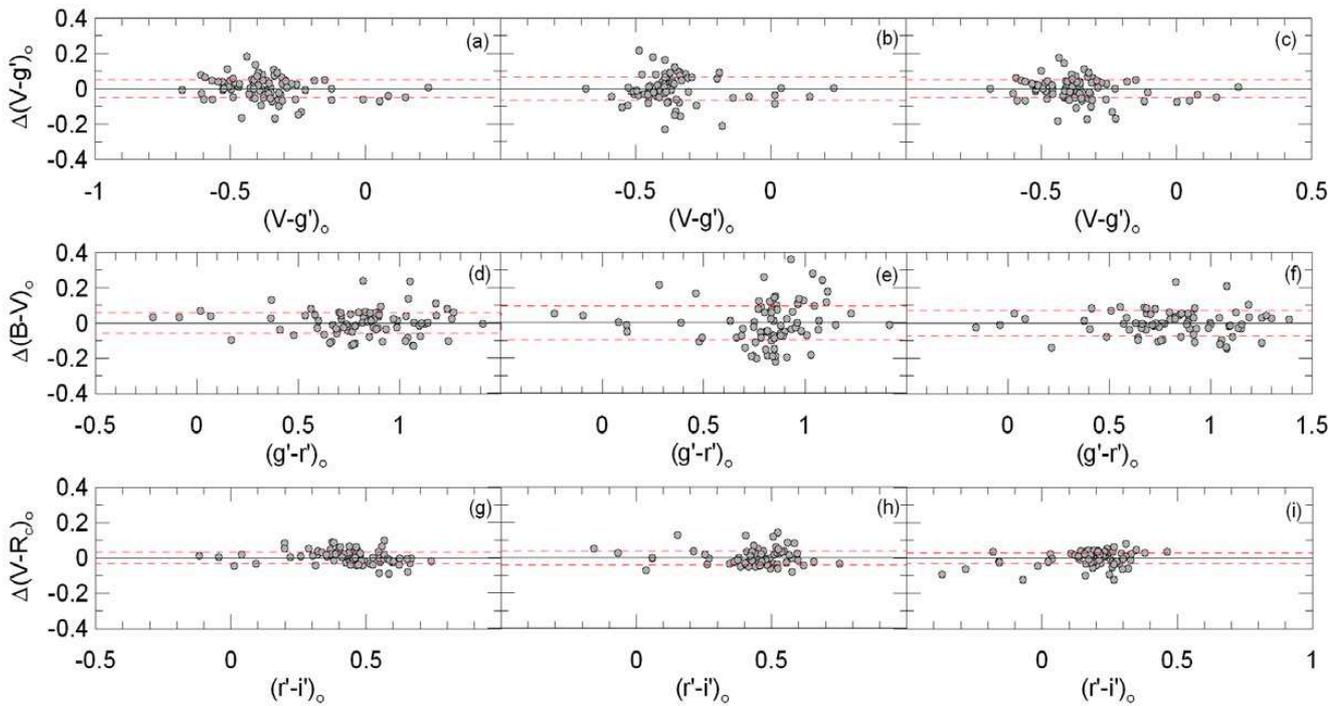}
\caption[] {Distributions of the residuals of the sample stars for 
inverse transformation equations (Eqs. 18-26).}
\end{center}
\end{figure*}

\begin{figure*}
\begin{center}
\includegraphics[scale=0.60, angle=0]{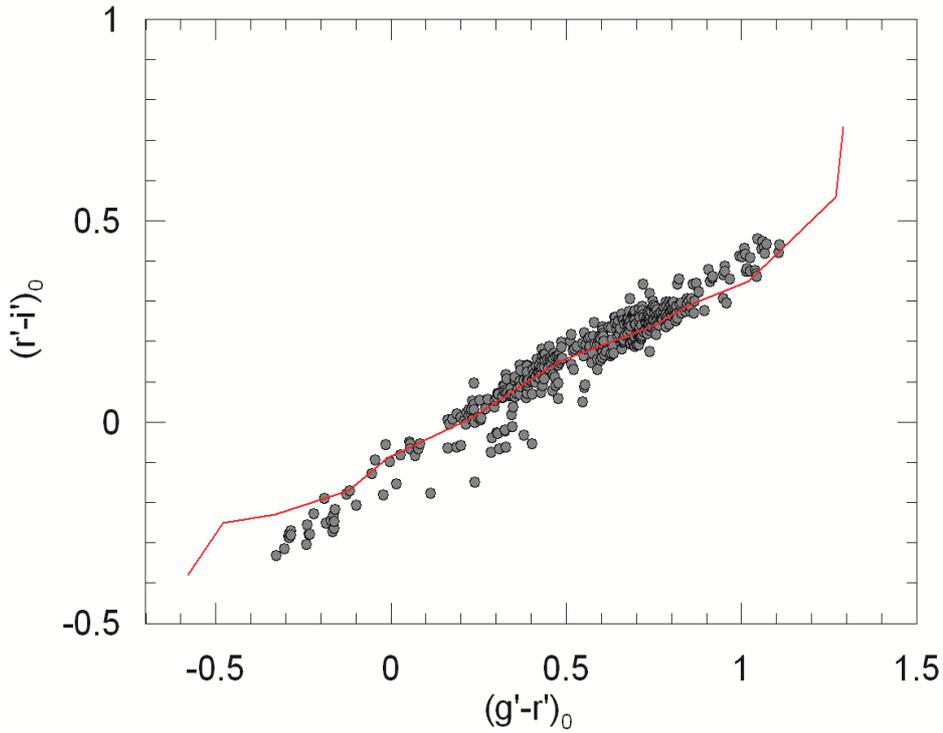}
\caption[] {$(g'-r')_0\times(r'-i')_0$ two-colour diagram of 427 stars 
used for the application of the transformation equations. The solid 
line indicates the synthetic two-colour diagram of \citet{Covey07}.}
\end{center}
\end{figure*}

\begin{figure*}
\begin{center}
\includegraphics[scale=0.85, angle=0]{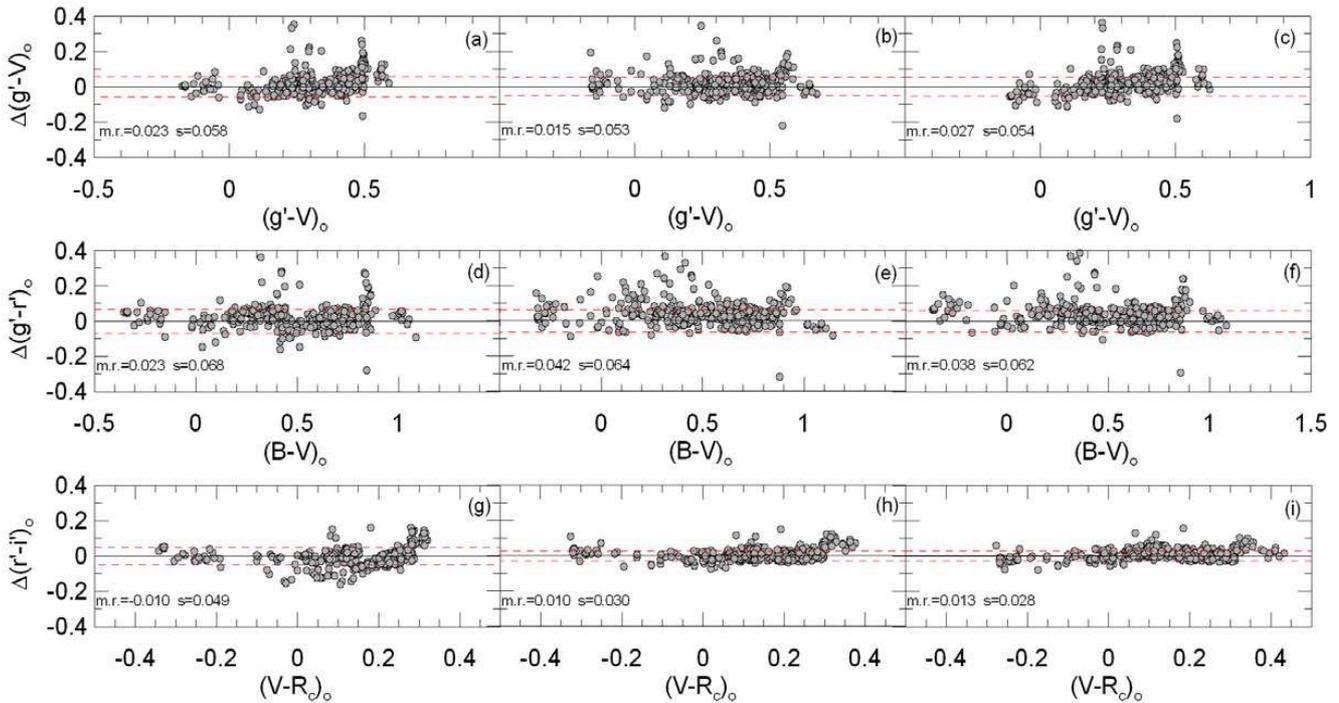}
\caption[] {Distributions of the residuals of 427 stars used for the 
application of the transformation equations. Mean residuals ($m.r.$) 
and standard deviations ($s$) are also indicated in each panel.}
\end{center}
\end{figure*}

\end{document}